\documentclass[pre,twocolumn]{revtex4} 
\usepackage{amsmath,epsfig,amssymb}

\newcommand{\be}{\begin{equation}}
\newcommand{\ee}{\end{equation}}
\newcommand{\bp}{\begin{picture}}
\newcommand{\ep}{\end{picture}}
\newcommand{\ba}[1]{\begin{array}{#1}}
\newcommand{\ea}{\end{array}}
\newcommand{\bea}{\begin{eqnarray}}
\newcommand{\eea}{\end{eqnarray}}

\begin{document}

\title{The Role of the Molecular Association in the Change of the Sign
  of the Soret Coefficient in Aqueous Mixtures}
\author{Bernard Rousseau}
\email{bernard.rousseau@lcp.u-psud.fr}
\affiliation{Laboratoire de Chimie-Physique, B\^{a}timent 350,
  Universit\'e Paris-Sud,91405 Orsay Cedex, France}
\author{Carlos Nieto-Draghi}
\author{Josep Bonet \'Avalos}
\affiliation{Departament d'Enginyeria Qu\'{\i}mica, ETSEQ, Universitat
  Rovira i Virgili, Avda. dels Pa\"{\i}sos Catalans 26, 43007 Tarragona Spain}
\date{\today}

\begin{abstract}
	The change of sign with composition in aqueous mixtures of
  associating fluids is analysed by means of Molecular Dynamics
  simulations. The results obtained are in quantitative agreement with
  the experimental data in water-ethanol and water-methanol solutions,
  which exhibit the mentioned change of sign. A subsequent theoretical
  analysis is addressed to establish a relationship between the
  dependence of the Soret coefficient with composition with the
  existence of large inter-species interactions. Although the change
  of sign of the Soret coefficient with composition is not only due to
  so-called chemical contribution analysed here, we discuss the role
  that these interactions play in such a change of sign.
\end{abstract}

\maketitle

Aqueous solutions often exhibit a non-ideal behaviour in both
thermodynamic and in dynamic properties, due to the existence of
strong, very directional hydrogen-bond interactions~\cite{mishima98,
errington}. One of the most astonishing features observed in many
aqueous mixtures is the change of the sign of the {\em thermal
diffusion} factor, or Ludwig-Soret coefficient, with composition
~\cite{Tyrrell-1961}.  This phenomenon has been observed since long
time ago but despite many efforts there is not even a qualitative
microscopic picture of the effect yet. The work addressed in this
letter is twofold: on one hand, we present, to the best of our
knowledge, the first molecular dynamics simulations able to
quantitatively reproduce the experimental values of the thermal
diffusion factor in aqueous solutions of associating fluids, including
the change of sign of this coefficient with composition. In the light
of these results and MD simulations of mixtures of Lennard-Jones
particules of hypothetical compounds, we analyse the role played by
strong cross-interactions on the composition dependence of the thermal
diffusion coefficient, and its influence on its change of sign.

When a binary mixture is under a thermal gradient, one
component will enrich in the cold region while the other will migrate
towards the hot boundary. At the stationnary state, the magnitude of
the separation is characterised by the thermal diffusion factor
\begin{equation}
\alpha_T \equiv - \frac{T}{x_A(1-x_A)}\;\frac{\nabla x_A}{\nabla T}
\;,\label{1} 
\end{equation}
where $x_A$ is the molar fraction of species $A$. The Soret
coefficient, $S_T = \alpha_T/T$ is also commonly used to quantify the
amplitude of the separation. The sign of $\alpha_T$ has a very
important meaning: it reflects the direction of separation of the
components of the mixture when a thermal gradient is applied. The
generally adopted convention is to take component $A$ as the heaviest
one. Thus, $\alpha_T$ is positive when the heaviest species ($A$)
enrich at the cold side, {\em i.e.} $\nabla x_A$ and $\nabla T$ have
opposite sign. In gas mixtures, $\alpha_T$ is proportional to the mass
difference between species, is positive, although this depends on the
interaction potential~\cite{Furry-1939}, and mostly independent of
composition. The situation is less simple in dense fluids. In a recent
paper, Debuschewitz and K\"ohler~\cite{Debuschewitz-2001} have shown
that the Soret coefficient can be splitted into independent
contributions: a mass effect, depending on the mass and moment of
inertia difference of the species, and a so-called {\em chemical}
effect. The mass effect is found to be independent of composition,
while the composition effect comes solely from the chemical part. In
benzene-cyclohexane isotopic mixtures, a change of sign was even
observed at a given composition. In mixtures of associating fluids, in
some ternary polymer solutions~\cite{Giglio-1977,deGans-2003} and
electrolyte solutions~\cite{Colombani-1998}, $\alpha_T$ strongly
depends on composition. Therefore, the composition dependence of the
thermal diffusion factor seems to be completely controled by the
different molecular interactions between the constituants, which has
been referred to as chemical effect.

To analyse this problem in more depth, we have performed molecular
dynamics simulations on several mixtures where strong interactions
exist. We present studies of water-ethanol and water-methanol mixtures
for which experimental data are available. Moreover, we have also
investigated water-acetone and water-dimethyl-sulfoxide mixtures, for
which we give our predictions. For all studied systems, a strong
composition dependence and a change of sign is observed.

Previous applications of MD simulations to the study of thermal
diffusion have been mostly devoted to the development of
methodologies~\cite{MacGowan-1986,Vogelsang-1987,Hafskjold-1993} and
to the study of Lennard-Jones like mixtures~\cite{Reith-2000,
Hafskjold94}.  There have been only a few attempts to compute the
thermal diffusion factor in molecular liquids~\cite{schaink93,
simon98}. Recently, however, the ability of MD simulations in
quantitatively reproducing thermal diffusion in pentane--{\em
n}-decane mixtures has been confirmed~\cite{perronace02b}.

%
\begin{figure}[ht]
\includegraphics[width=5.5cm,angle=270]{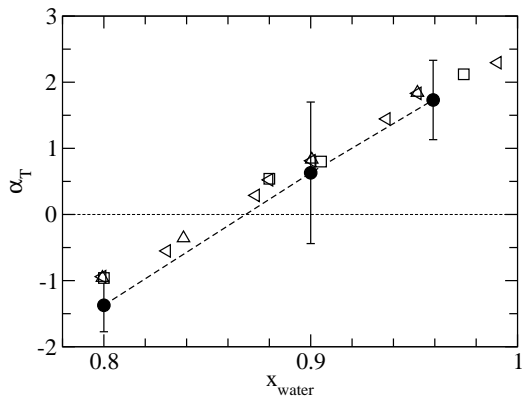}
\caption{Thermal diffusion factor versus water molar
  fraction in water-ethanol mixture at 298~K and 1~atm. Simulation
  data: {\large $\bullet$}, experimental data:
  $\square$~\cite{bou-ali99}, $\vartriangleleft$~\cite{Kolodner-1988},
  $\triangle$~\cite{Zhang-1996}. Dashed line is a guide to the eyes.
\label{alpha-w-etoh}}
\end{figure}

In our simulations we have used a non-equilibrium
algorithm~\cite{nieto03b,Muller-Plathe-1999}, which permits to
maintain a heat flow through the system with control on the average
temperature and total momentum and energy conservation. As far as the
molecular potentials are concerned, we have chosen simple models able
to reproduce strong interactions such as hydrogen bonds. The following
force fields were used for self-interactions: for water, the TIP4P
model~\cite{jorgensen83}; for methanol and ethanol, the OPLS model of
Jorgensen~\cite{jorgensen86}; for acetone, a model proposed by Wheeler
and Rowley~\cite{wheeler}; and, finally, for DMSO, the force field
derived by Luzar and Chandler~\cite{luzar93}.  Standard
Lorentz-Berthelot mixing rules were used for cross interactions. The
simulations were carried out on 800 molecules in water-alcohol systems
and 500 in the other cases. The details of the simulations will be
given elsewhere.

The simulation results for the water-ethanol mixture at ambient
conditions are compared with experimental data on
fig.~\ref{alpha-w-etoh}. It is important to notice that experimental
data have been obtained by three different
methods~\cite{bou-ali99,Kolodner-1988,Zhang-1996}. The consistency
between experimental results is remarkable and these data can be
considered as reference data for this system. As can be seen, our
simulation results are in very good agreement with experiments. This
validates our methodological choices. A change of sign of $\alpha_T$
is observed for $x_{\mbox{water}} \approx 0.86$.

\begin{figure}[ht]
\includegraphics[width=5.5cm,angle=270]{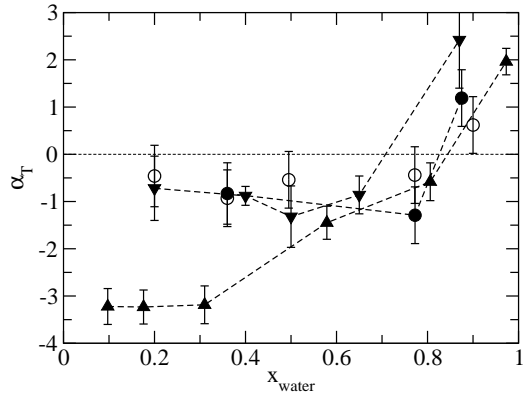}
\caption{Thermal diffusion factor versus water molar
  fraction in water-methanol at 313~K and 1~atm: NEMD data~({\Large
    $\bullet$}) and experiments~({\Large $\circ$}) from Tichacheck {\em et
    al.}~\cite{Tichacek-1956}.  NEMD data for water-DMSO
  ($\blacktriangledown$) and water-acetone ($\blacktriangle$) at 298~K
  and 1~atm. Dashed lines are guide to the eyes.
\label{alpha-w-ac_dmso_met}}
\end{figure}

In fig.~\ref{alpha-w-ac_dmso_met}, we compare simulation data with
experiments for the water-methanol system at 313~K and 1~atm.
Experimental data are those of Tichacek {\em et
  al.}~\cite{Tichacek-1956} at the same thermodynamic conditions. We
observe a plateau at low water molar fraction followed by a change of
sign of $\alpha_T$ at $x_{\mbox{water}} \approx 0.8$, in agreement
with the experimental data.  Finally, predictions for the thermal
diffusion factor versus water content in water-acetone and
water-dimethyl-sulfoxide mixtures are shown in
fig.~\ref{alpha-w-ac_dmso_met}. In both cases, a change of sign of
$\alpha_T$ is also observed. Therefore, an important dependence of
$S_T$ with composition is observed in these mixtures having strong
hydrogen-bonds interactions. 
%
%

Despite the ability of MD simulations in reproducing the experimental
behaviour, the question about what microscopically causes the strong
composition-dependence and, eventually, the change of
sign, in $\alpha_T$ remains still open. To shed some light on this
question we have built a two-dimensional Ising model in a temperature
gradient. In this model, the dynamic process consists of the swap of 2
particles between neighboring sites $i$ and $j$, kept at different
local temperatures $T_i$ and $T_j$, according to an externally fixed
temperature gradient. The presence of species A (or B) on a given
lattice site $i$ is represented by the state $\vec{\sigma_i}=(1,0)$ (or
$(0,1)$) of the local spin.  The site-site interactions are limited to
first nearest neighbors, and the lattice has no vacancies.  The
Hamiltonian is given by $H=\sum_{i,j>i} \vec{\sigma}_i\cdot
{\mathbf{J}} \cdot \vec{\sigma}_j$, where $\mathbf{J}$ is a
constant square matrix which defines the particle-particle
interactions through its elements: $\varepsilon_{AA}$,
$\varepsilon_{BB}$, for the intra-species interactions, and
$\varepsilon_{AB}=\varepsilon_{BA}$ for the inter-species. The
necessary particle number conservation in the process implies a
spin-exchange or Kawasaki dynamics~\cite{kawa66a}. The actual dynamic
process is determined through the transition probability for one
spin-exchange, which has been chosen to be
\begin{eqnarray}
\lefteqn{W([\vec{\sigma_i},\vec{\sigma_j}] \rightarrow
  [\vec{\sigma_i'},\vec{\sigma_j'}])=} 
 \label{transition} \\ 
 & & \mbox{min}\left[ e^{ - \left[ \left(E_i'-E_i\right)/kT_i +
 \left(E_j'-E_j\right)/kT_j \right] },1 \right] \nonumber 
\end{eqnarray} 
In this expression, $E_i$ is the energy of the spin at the site $i$,
that is
\begin{equation}
E_i=\sum_{k\in(nn)}\vec{\sigma}_i\cdot \mathbf{J} \cdot \vec{\sigma}_k
\label{eki}
\end{equation}
where $nn$ stands for {\em nearest neighbors}. This transition
probability satisfies the required detailed balance condition so that
the appropriate thermodynamic equilibrium is reached if the
temperature is uniform. 

\begin{figure}[ht]
\includegraphics[width=5.5cm,angle=270]{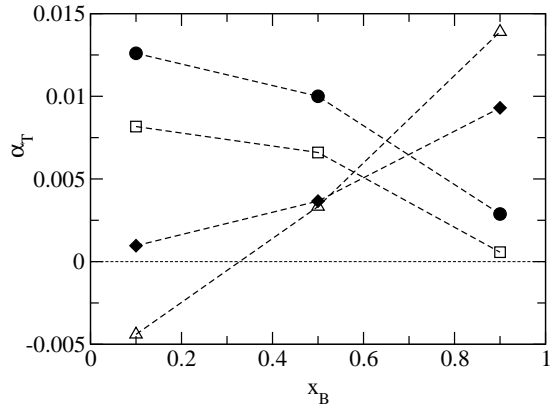}
\caption{Thermal diffusion factor $\alpha_T$ for the mixture
  of particles in function of the molar fraction of one component for
  several cross interaction parameters
  $\varepsilon_{AB}= -1.0$ ({\Large $\bullet$}), $-1.414$
  ($\square$), $-2.0$ ($\blacklozenge$), $-2.546$ ($\vartriangle$). Dashed
  lines are guide to the eyes.\label{lattice_c1}}
\end{figure}

In fig.~\ref{lattice_c1} we present the results of simulations on the
lattice model for different values of $\varepsilon_{AB}$, with
$\varepsilon_{AA}=-2.0$ and $\varepsilon_{BB}=-1.0$ in dimensionless
units, where $T^* = 2.55$ is the average reduced temperature.
%
The figure clearly shows that the slope of the concentration
dependence of $\alpha_T$ changes from positive to negative when
$\varepsilon_{AB}$ varies from $\varepsilon_{BB}$ to
$\varepsilon_{AA}$. When
$|\varepsilon_{AB}|>|\varepsilon_{AA}|,|\varepsilon_{BB}|$, a change
of sign of $\alpha_T$ with composition is in addition observed. This
is again a strong suggestion that in dense systems the behaviour of
$\alpha_T$ with composition is severely influenced by interations
between unlike species relative to interactions between like species.

Obviously, the lattice model is too simplistic to draw conclusions,
but it grasps an important aspect of the phenomenon. It can be
therefore interesting to evaluate the role played by the unlike
species interactions by doing direct molecular dynamics simulations on
a simple Lennard-Jones (LJ) mixture, which corresponds to a more
realistic situation. We have thus computed $\alpha_T$ using the same
MD methodology as previously described. Both species have the same
mass and size ($\sigma=3.405$~\AA), but different energetic parameters
($\varepsilon_{AA} = 0.996$, $\varepsilon_{BB} = 1.384$~kJ~mol$^{-1}$.
Notice that the values for particle $A$ correspond to argon). For the
inter-species energetic parameter, we have used a value of
$\varepsilon_{AB}$ = 2.348~kJ~mol$^{-1}$, much larger than parameters
between like species. The thermal energy is $RT =0.998$~kJ~mol$^{-1}$.
Simulations have been done in the dense liquid state and care have
been taken that no demixion occured. A change of sign of $\alpha_T$ is
again observed. Effectively, for the following compositions, $x_B=0.1,
0.5, 0.9$, the values of the thermal diffusion are, respectively,
-1.97, -0.187, +1.679. The fact that systems of Lennard-Jones
particles could exhibit a change of sign of $\alpha_T$ with
composition has remained unnoticed in previous works, partly because
most of studies have made use of the standard Lorentz-Berthelot mixing
rule, which place $\varepsilon_{AB}$ between $\varepsilon_{AA}$ and
$\varepsilon_{BB}$. Obviously, the interesting physics underlying the
behaviour of these simple systems under a temperature gradient
deserves a deeper analysis and work is in progress.

All these observations therefore suggest that in more realistic
systems the chemical potentials at a given
space point $z$ are analogous to the energy $E_i$ of a given particle
at the $i^{th}$ site in the lattice model, in view of
eq.~(\ref{eki}). One may thus think the physical system as divided in
small mesoscopic slices and the dynamics of matter transfer as being
described by a Master equation. In analogy with the Ising model
described above, one may argue that the transition probability for an
exchange of $\delta n_A$ moles of $A$ at the slice located at $z$ by
$\gamma \delta n_B$ moles of $B$, initially at the slice $z+\delta$ is
proportional to
\begin{equation}
W[z, z+\delta]  \propto 
\frac{e^{ - \left(\mu_A(z)-\gamma \mu_B(z)\right)/kT(z)}}{ e^{-\left(
      \mu_A(z+\delta)-\gamma \mu_B(z+\delta)\right)/kT(z+\delta)}}
\label{gentrans} 
\end{equation}
The proportionality factor $\gamma$ is introduced due to the fact that
the matter exchange is not necessarily equimolar. We have considered
that the variation of volume in the switch of $\delta n_A$ moles by
$\delta n_B$ should be negligible, so that $\overline{v}_A \delta n_A
+ \overline{v}_B \delta n_B =0$ and hence $\gamma =
\overline{v}_A/\overline{v}_B$. In eq.~(\ref{gentrans}) $\delta$
should be considered as a characteristic microscopic length analogous
to the mean free path in kinetic theory of gases.  Respectively,
$\mu_A$ and $\overline{v}_A$ together with $\mu_B$ and
$\overline{v}_B$ are the local chemical potentials and partial molar
volume of species $A$ and $B$. In steady state, $\partial n_A/\partial
t=0$ and thus a probability balance indicates that
\begin{eqnarray}
\lefteqn{\left\{n_A(z+\delta)W[z+\delta , z]+n_A(z-\delta)W[z-\delta ,
    z] \right.} \label{master} \\ 
&& \left. -n_A(z) \left(W[z, z-\delta]+W[z, z+\delta] \right)
\right\} =0     \nonumber \;.
\end{eqnarray}
Taking $\delta$ as a length much smaller than the spatial variation of
the thermodynamic fields $T(z)$ and $n_A(z)$, an expansion in powers
of $\delta$ reveals that, in a system in steady state with no net
matter flow, the condition
\begin{equation}
\nabla \left(\frac{\mu_A-\gamma \mu_B}{kT}\right)_P =0 \label{cond}
\end{equation}
should be satisfied. In the spirit of Onsager's linear irreversible
thermodynamics~\cite{degroot}, this result indicates that the
effective thermodynamic force causing the whole effect of species
separation in a liquid in a temperature gradient is the left hand side
of (\ref{cond}). Thus, eq.~(\ref{cond}) states that in a system
without matter flow, the variation of $(\mu_A-\gamma \mu_B)/kT$ with
temperature is compensated by the variation of the same quantity with
composition. This point of view permits to obtain a relationship
between the temperature gradient and the composition gradient and,
therefore, of the thermal diffusion factor, according to
eq.~(\ref{1}),
\begin{eqnarray}
\alpha_T =- \frac{\overline{v}_A\overline{v}_B}%
                 {x\overline{v}_A+(1-x)\overline{v}_B}
            \frac{\overline{h}_A/\overline{v}_A-\overline{h}_B/\overline{v}_B}%
                 {x\left(\partial \mu_A/\partial x \right)_{P,T}} \;,
                 \label{17}  
\end{eqnarray}
where $\overline{h}_A$ and $\overline{h}_B$ are the partial molar
enthalpies of the components $A$ and $B$, respectively. In the
derivation, $\gamma$ has been taken as constant. Surprisingly, this
expression is identical to that proposed by Kempers and very close to
that of many authors that have used thermodynamic considerations
(see~\cite{kempers89} and references therein). The accuracy of this
expression has been tested on hydrocarbon mixtures~\cite{kempers01},
with a reasonable agreement. In associating fluids it is thus expected
a change in the sign of the thermal diffusion factor with composition
if $\overline{h}_A/\overline{v}_A-\overline{h}_B/\overline{v}_B$
changes sign at some particular point. This change of sign in our
picture is caused by large inter-species interactions, larger than the
intra-species interactions, which makes the dilute component to be
more strongly bound than the concentrated component and then to have a
more negative chemical potential. If the molar volumes are the same
and, for instance, $A$ is the dilute component, any composition
fluctuations will favor $A$ to go downhill in temperature and
accumulate preferably on the cold side. This effect is reversed if we
vary the concentration up to the point where $B$ is the more dilute
component: $B$ is then more strongly bound than $A$ and any
composition fluctuations will favor an accumulation of $B$ on the cold
side of the box, thus reversing the sign of the Soret coefficient. If
the molar volumes are different, any composition fluctuation involves
different number of particles of both species present, to maintain the
volume constant.  Then it is essentially the value of the partial
molar enthalpy per unit volume of the components,
$\overline{h}_{A,B}/\overline{v}_{A,B}$, which controls the sign of
the separation, as it has been already been pointed out in the
literature~\cite{Prigogine-1950}.

Equation~(\ref{cond}) can be derived maximizing the appropriate Legendre
transform of the entropy, the so-called Massieu-Planck function
$S[1/T,P/T]$~\cite{callen85}, locally defined, and subject to the
constraint of the constant volume in the composition exchange. This is
in fact the approach followed by Kempers. Under these conditions, the
separation of the components is purely due to thermodynamic effects,
since $(\mu_A-\gamma \mu_B)/k_B T$ is constant along the
sample. However, the theory of irreversible processes also considers
the existence of a contribution to the thermal diffusion due to purely
kinetic origin~\cite{degroot}, which is completely ignored in the
hypothesis underlying eqs.~(\ref{transition}) and~(\ref{gentrans}). An
explicit consideration of this kinetic effect would imply the right
hand side of eq.~(\ref{cond}) to be non-zero.  Incorporating such a
kinetic effects to the thermodynamic contributions related to the
partial molar enthalpies is, to our believe, the main goal of a
predictive theory for the Soret effect in describing both, chemical
and kinetic effects, in dense systems as well as in dilute gases.

\begin{acknowledgments}
  This work was partly supported by the Spanish Government {\it
    Ministerio de Educaci\'{o}n Ciencia y Tecnolog\'{i}a} (grants No.
  PPQ2000-2888-E, PPQ 2001-0671), {\it Generalitat de Catalunya}
  (project ACI2001 and ACI2002) and the URV (project No. 2000PIR-21).
  C.~Nieto-Draghi thanks Universitat Rovira i Virgili (Spain) and {\it
    Ministerio de Educacion Cultura y Deporte} (Spain) for financial
  support. 
\end{acknowledgments}

\bibliography{soret_cond_mat}

\end{document}